\documentclass[prb,preprint,amsmath,amssymb]{revtex4}

\usepackage[dvips]{color}
\usepackage{graphicx} 

\usepackage{amsmath}

\newcommand{\beq}{\begin{equation}}
\newcommand{\enq}{\end{equation}}

\newcommand{\dsee}{{\delta\sigma^{ee}}}
\newcommand{\dswl}{{\delta\sigma_{}^{wl}}}
\newcommand{\dsgeo}{{\delta\sigma_{}^{G}}}

\newcommand{\Squatre}{S1}
\newcommand{\Sdeux}{S2}

\begin{document}
\title{Interplay between interferences and electron-electron interactions in epitaxial graphene}
\author{B. Jouault$^{1,2}$,  B. Jabakhanji$^{1}$, N. Camara$^{1,3}$, W. Desrat$^{1}$, 
        C. Consejo$^{1}$, J. Camassel$^{1,2}$}        
        \affiliation{$^1$Universit\'e Montpellier 2, Groupe d'\'Etude des
  Semiconducteurs, cc074,  pl. Eug\`ene Bataillon, 34095 Montpellier cedex 5, France}
\affiliation{$^2$ CNRS, UMR 5650, cc074,  pl. Eug\`ene Bataillon, 34095 Montpellier cedex 5, France}
\affiliation{$^3$ CNM, Campus UAB, Bellaterra (Barcelona), Spain}
\begin{abstract}
We separate localization and interaction effects in epitaxial graphene devices grown on the C-face of an 8-$^o$off 4H-SiC substrate  by analyzing the low temperature conductivities. Weak localization and antilocalization are extracted at low magnetic fields, after elimination of a geometric magnetoresistance and subtraction of the magnetic field dependent Drude conductivity. The electron electron interaction correction is extracted at higher magnetic fields, where localization effects disappear. Both phenomena are weak but sizable and of the same order of magnitude.
    If compared to graphene on silicon dioxide, electron electron interaction on epitaxial graphene are not significantly reduced by the larger dielectric constant of the SiC substrate. 
\end{abstract}

\maketitle
\section{introduction}

Graphene-based devices are exciting candidates for future generations of microelectronic devices. 
A promising technique to produce graphene  at an industrial scale  is the epitaxial graphene growth from a SiC substrate,
because these SiC substrates can be patterned using standard lithography methods.
Thanks to recent technical improvements, the most delicate and intrinsic features of graphene,
those reflecting the chiral nature of the quasiparticles, as, for instance,
the so-called 'Half Integer quantum Hall effect'~\cite{shen:172105,jobst-2009,laraavila-2009,wu-2009-95} and the weak antilocalization~\cite{berger2007}, have been recently reported for epitaxial graphene.
    
In  graphene, depending on the relative magnitude of intervalley scattering time 
and phase coherence time,
either weak localization (WL) or weak antilocalization (WAL) has been predicted~\cite{kechedzhi2007}. 
The phase interference correction to the resistance depends on the nature of the disorder. 
For epitaxial graphene samples, 
elastic scattering favorable for WAL can be caused by remote charges like ionized impurities in the substrate.
On the other hand, atomically sharp disorder (local defects, edges)
causes intervalley scattering and gives rise to WL. 

Experimentally, in two-dimensional gases, WL is often mixed with 
electron-electron interaction~\cite{altshuler80} (EEI).
As WL, EEI gives a correction to the Drude conductivity with a $\ln T$ dependence.
It follows that the experimental extraction and  separation of  EEI and WAL contributions are usually difficult~\cite{goh2008}. 
For graphene, in the diffusive regime, EEI is expected to
give this usual temperature dependence correction proportional to $\ln(T)$. 
EEI is also expected to be sensitive to the different kinds of disorders~\cite{savchenko2010}. 

In this work, we take advantage of  simultaneous measurements of the longitudinal and transverse resistances 
to invert the resistivity tensor.
We can then separate the different mechanisms which give corrections to the Drude conductivity: 
i) a geometric contribution, which experimentally appears as a constant term in the longitudinal conductivity;
ii)  WL and WAL, by a comparison between the experimental magnetoconductance and the Drude magnetoconductivity,   
iii) EEI, whose temperature dependence can be analyzed in the magnetic field range for which WAL has disappeared.

\section{Experimental details}

We have used large and homogeneous single epitaxial graphene layers grown on the C-face of an insulating 8$^o$ off-axis 4H-SiC substrate. The graphene sheets have an elongated triangular shape of which quality and homogeneity can be easily checked using micro-Raman spectroscopy. 
On few selected samples, Cr/Au ohmic contacts were deposited to define Hall-bars with a rough geometry (see inset in Fig.1 for details). 
Then magneto-transport measurements were done using a 14T magnet in a cryostat and a Variable Temperature Inset  operated down to 1.5K. 
On the best samples, at high magnetic field, the half-integer quantum Hall effect could be observed up to the last plateau in the temperature range 1.5 to 40K. 
For details, see Ref.~\cite{camaraEHQ}.  
In this work we focus on two moderately doped samples 
($\Squatre$ and $\Sdeux$) with dimensions, carrier concentration, mobility and scattering times reported in table~\ref{table1}.
We work at low injection currents (10nA- 1$\mu$A), mainly at low magnetic fields ($|B| \le$ 3T), low temperatures (1.5 K-200K) and we focus on the magnetic field dependence 
of the longitudinal and transverse elements of the resistivity tensor.

\begin{table}[b!]
\begin{tabular}{|l|c c c c c c cc |}
\hline
sample  & $L$  & $W$   & $n_s$ & $\mu$ & $\tau_{sr}$ & $\tau_{lr}$ & $\tau_{tr} $& $D$ \\
\hline
$\Squatre$     &  5    &  5  &  1.1   & 5000 &    0.57    &  0.1  &  0.065   & 0.03         \\ 
$\Sdeux$     &  40   &  10  & 0.8   & 11000  &  0.48    &  0.15 &  0.11   &   0.057    \\     
\hline
\end{tabular}
\caption{for sample $\Squatre$ and $\Sdeux$: length $L$ between lateral probes, width $W$ (in $\mu$m), hole concentration (in 10$^{12}$cm$^{-2}$), mobility (in cm$^2$V$^{-1}$s$^{-1}$), scattering times $\tau_{sr}$, $\tau_{lr}$ and $\tau_{tr}$ (in ps) and diffusion constant (in m$^2$s$^{-1}$).}
\label{table1}
\end{table}

\section{Background Considerations}

\begin{figure}
\includegraphics[width=0.9 \columnwidth]{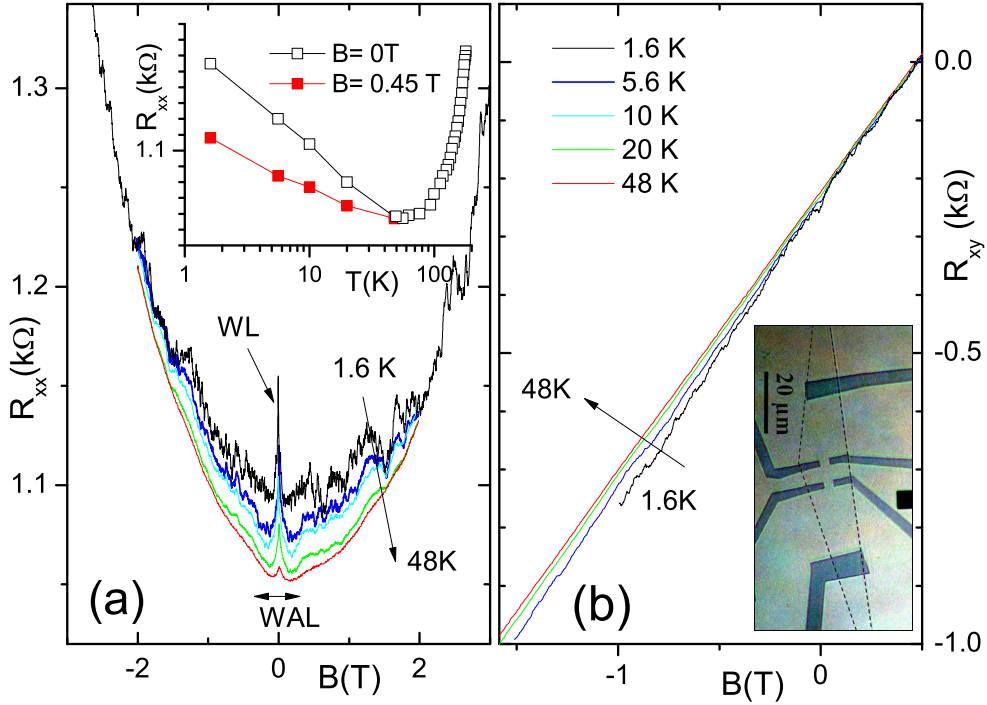}
\caption{(color online) (a) $R_{xx}(B)$ at different temperatures for sample $\Squatre$. All curves crosses at $B \approx 2 T$ and have a $B^2$ behavior. Inset shows the temperature dependence of $R_{xx}$ at $B$= 0 T (black open squares) and $B$= 0.45 T (solid red squares).
         (b) $R_{xy}(B)$ at different temperatures. The slope of the Hall voltage increases when $T$ decreases due to electron-electron interactions. Inset: photograph of sample $\Squatre$, edges of the graphene layer have been indicated by a dashed line.}
\label{fig:rhrl}
\end{figure}
An overview of the results obtained for sample S1 is shown in Fig.1.  
The experimental longitudinal and transverse magnetoresistances $R_{xx}(B)$ and $R_{xy}(B)$ for sample $\Squatre$  are presented in Fig.~1a and 1b respectively, at different temperatures between 1.6K and 48K. 
Let us consider first, Fig. 1a.
At low fields ($B \le $ 0.1 T), a negative magnetoresistance peak centered at $B=0$T is observed in $R_{xx}$.
This peak is typical of weak localization (WL). Indeed, this peak a linear dependence of $R_{xx}$ {\it vs} $\ln(T)$ with a slope of the order of $h/e^2$ (see inset of Fig.~1a) and the amplitude of the peak is $\Delta \rho_{xx} \approx 2 \rho_{xx}^2 e^2/h$ ($\rho_{xx}$ is the longitudinal resistivity). At higher magnetic fields, reproducible fluctuations in the conductance $G$ are also observed, with 
amplitudes 
$\Delta G \approx 0.5  e^2/h$ for $\Squatre$  
and $\Delta G \approx 0.1 e^2/h$ for $\Sdeux$.
 
At magnetic fields $|B| \le 0.4$T, a smooth depression is observed in $R_{xx}$, which we
attribute to weak antilocalization (WAL). Because this WAL is barely visible in Fig. 1a, we also present additional
experimental results in Appendix A. Experimentally, WAL is blurred because
fluctuations of conductance are also present, and also because
the positive magnetoresistance of the WAL is superposed to another  
positive magnetoresistance with a pronounced parabolic dependence,
which is well observed on the whole magnetic field range $|B| \le 3$T presented in Fig.~1a.
The theory of weak localization is based on the diffusion approximation and does not hold
for magnetic fields much higher than $B_{tr}$= $\hbar/4eD\tau_{tr} \approx$ 100 mT,
where $\tau_{tr}$ is the transport relaxation time and $D$ the diffusion coefficient.
This means that the parabolic background above $B_{tr}$ should not be attributed to WL features. It must have a different origin. 
Since the device is being far from having the shape of an ideal Hall-bar (see inset in Fig.1b) 
we ascribe this parabolic magnetoresistance component to  magnetic deflection of the current lines.
This is because the lateral probes are invasive 
and because the graphene layer under the lateral contacts is likely to have  different mobility and carrier concentration~\cite{contacts}. For details, see Annex B.

Beyond geometric and interference corrections, 
a third correction to the resistivity manifests.
It comes from electron-electron interactions and shows as:
i) the persistence of a $\ln(T)$ linear temperature dependence of $R_{xx}$, in a magnetic field range where
interference effects are suppressed (see inset of Fig. 1a for a field of $B=0.45 T$);
ii) a variation of the Hall slope (Fig. 1b) as a function of $T$, which cannot be explained by
 variations of the carrier density, as the hole gas is strongly degenerate ($E_F/k_B T \ge$ 20 for $T \le 50$K, where $E_F$ is the Fermi energy);
iii) a crossing of all $R_{xx}(B)$ resistances taken at different temperatures at $B \approx 1/\mu$ (Fig. 1a)~\cite{minkov2001}.

The justification of the last two points, originally given in~\cite{minkov2001},  is as follows.
EEI give no correction to the transverse conductance $\sigma_{xy}$.
It gives a small correction $\dsee$ to the longitudinal conductance $\sigma_{xx}$. This correction does not depend on $B$
if the magnetic field is smaller than a critical field $B_S \le \pi k_BT/2\mu_B $. This
condition if fulfilled for all the temperature and magnetic field ranges of this study.
Because of this correction, EEI gives rise to a negative magnetoresistance in the first order in $\dsee \ll \sigma_{xx}$:
\begin{equation}
\rho_{xx} \sim 1/\sigma_0  - (1- \mu^2 B^2 ) \dsee/\sigma_0^2
\end{equation}
and a variation of the Hall slope 
$\delta \rho_{xy}/\rho_{xy} \approx -2 \dsee / \sigma_0$
where $\sigma_0$ is the conductivity at $B=0$.
These relations are derived by inverting the conductivity tensor:
\begin{eqnarray}
\rho_{xx}=  \frac{\sigma_{xx}}{\sigma_{xx}^2+\sigma_{xy}^2} \label{eqn:sxx}\\
\rho_{xy}=  \frac{\sigma_{xy}}{\sigma_{xx}^2+\sigma_{xy}^2} \label{eqn:sxy}~.
\end{eqnarray}

It has been established recently~\cite{goh2008} that the separation of EEI and WL~\cite{goh2008} is simplified if the magnetoconductivities $\sigma_{xy}(B)$ and $\sigma_{xx}(B)$ are used, rather than the resistivities. 
As already stressed, the conductivity of the graphene layer is different below the Cr/Au contacts and,
strictly speaking, inverting the resistivity tensor is incorrect because the device is inhomogeneous. 
However, experimentally, the geometric correction is small, appears only in the longitudinal 
magnetoresistance as a parabolic correction, and consequently 
geometric corrections to the longitudinal conductivity appear
as a $B$-independent shift $\dsgeo$.

\begin{figure}
\includegraphics[width=0.9 \columnwidth]{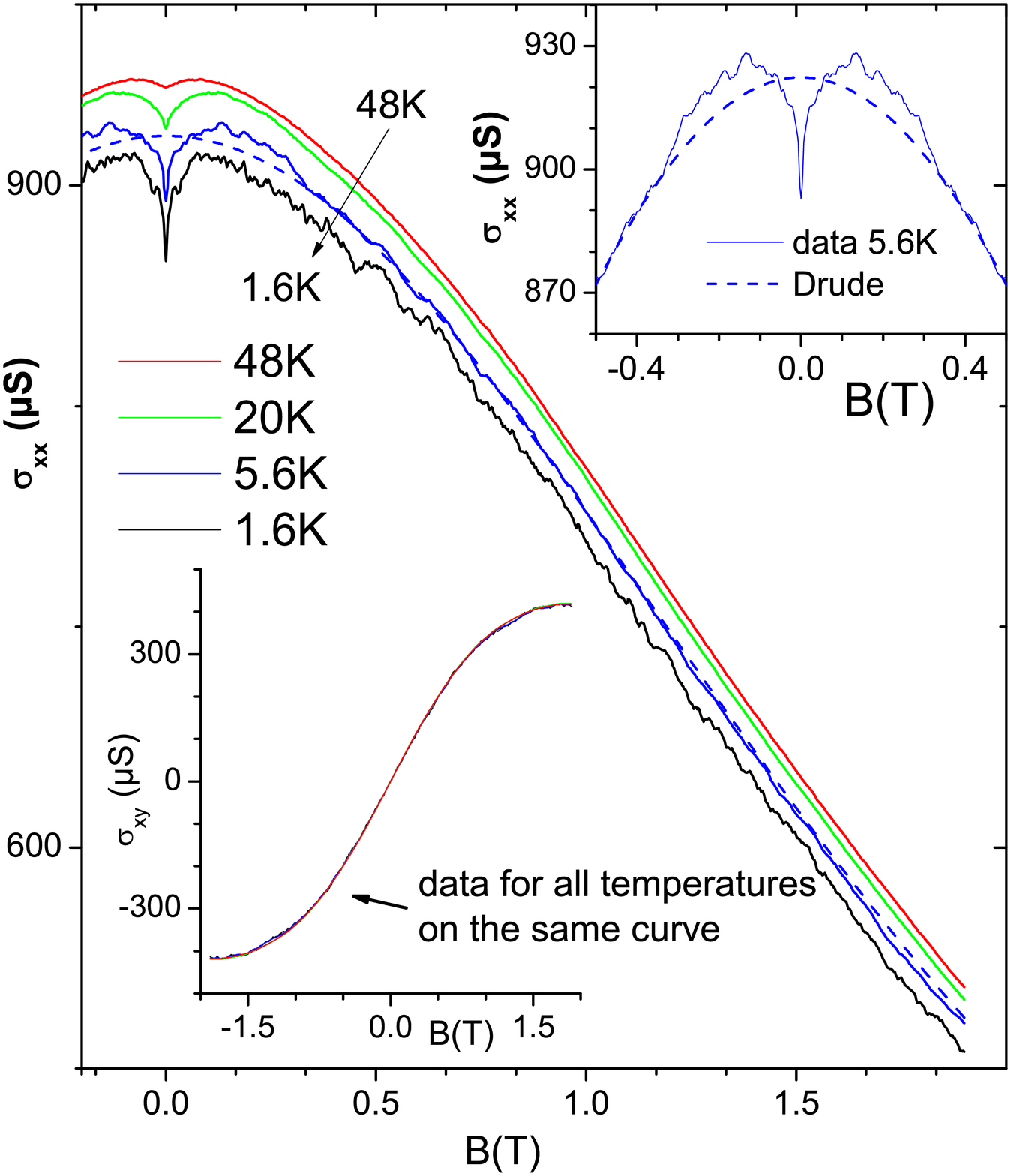}
\caption{(color online) longitudinal conductivity $\sigma_{xx}$ {\it versus} $B$ at different temperatures for sample $\Squatre$. Blue dashed line is the best fit of the conductivity  at $T$= 5.6K, on a $B$-interval 0.5-2 T. The fit is done according to $\sigma_{xx}^D(B)+ C$ (see text). The upper inset is the enlargement of the area around $B \sim 0$T. 
      The lower inset shows that all transverse conductivities  $\sigma_{xy}(B)$, taken  at different temperatures, collapse on a single curve.}
\label{fig:sxxsxy}
\end{figure}

The magnetoconductivities for sample $\Squatre$ are plotted in Fig.~\ref{fig:sxxsxy} at different temperatures.
The conductivities have been obtained from Eq.~\ref{eqn:sxx} and~\ref{eqn:sxy}, where the resistivities have been estimated by
$\rho_{xx} \approx (W/L) (R_{xx}(B)+R_{xx}(-B)) /2$ and
$\rho_{xy} \approx  (R_{xy}(B)-R_{xy}(-B)) /2$.
The lower inset of Fig.2 shows that all $\sigma_{xy}(B)$ taken at different temperatures collapse
on the same curve, as expected for EEI.
This also confirms that both mobility and carrier concentration are constant over the
whole temperature range and, 
as the geometric correction only depends on $T$ {via} the Hall angle $\mu B$, 
that $\dsgeo$ does not depend on the temperature. Finally, 
the $T$-independence of the transverse conductivity indicates an effective separation of
EEI  ($\dsee$) and  interference ($\dswl$) corrections over almost the whole $B$-range.
At low fields $|B| \le B_{tr}$, a $T$-dependence of $\sigma_{xy}$ due to WL is expected but
is not observed, being beyond our experimental resolution.

The large decrease of $\sigma_{xx}(B)$ between 0 and 2T in Fig.~\ref{fig:sxxsxy} is due 
to the magnetic field dependence of the Drude conductivity
$\sigma^D= \sigma_0 / (1+ \mu^2 B^2)$.
We express the total conductivity  by 
\begin{equation}
\sigma_{xx}= 
\sigma^D+ \dsee + \dsgeo + \dswl.
\label{eqsigma}
\end{equation}
When WL is negligible: $|B| \ge 0.5 \gg B_{tr}$, 
we impose $\dswl=0$ and Eq.~\ref{eqsigma}, for a given temperature,
simplifies as  $\sigma_{xx}=\sigma^D + C$, where $C$ is a constant which
incorporates both geometric and EEI corrections.
This last formula gives very good fits to the conductivity at all temperatures.
As an example, the blue dotted line in Fig.~2 is the fit for the  $T$=5.6K data in the interval 0.5-2 T,
where $\sigma_0$, $\mu$ and $C$ are the fit parameters.
Transverse and longitudinal resistivities can be fitted separately
but still give very similar mobilities and concentrations, 
which are reported in Table~I.

The upper inset in Fig.~2 is an enlargement of the low-$B$ data at $T=5.6K$ and
evidences the weak antilocalization peak which takes place
around the weak localization centered at $B=0$. A similar weak antilocalization can be detected for all temperatures.  
The WL theory does not take into account the modification of the current paths due to magnetic field. In other words, 
before examining the WL contribution to the conductivity, we should first subtract the Drude fits $\sigma^D+C$
(given by the dotted line in Fig.~\ref{fig:sxxsxy})
to the observed longitudinal conductivities~\cite{pros2001}. Experimental results for samples 
$\Sdeux$ and $\Squatre$ are shown in Fig.~\ref{fig:wl}, for different temperatures.

\section{Correction due to Weak localization and antilocalization}

\begin{figure}
\includegraphics[width=0.9 \columnwidth]{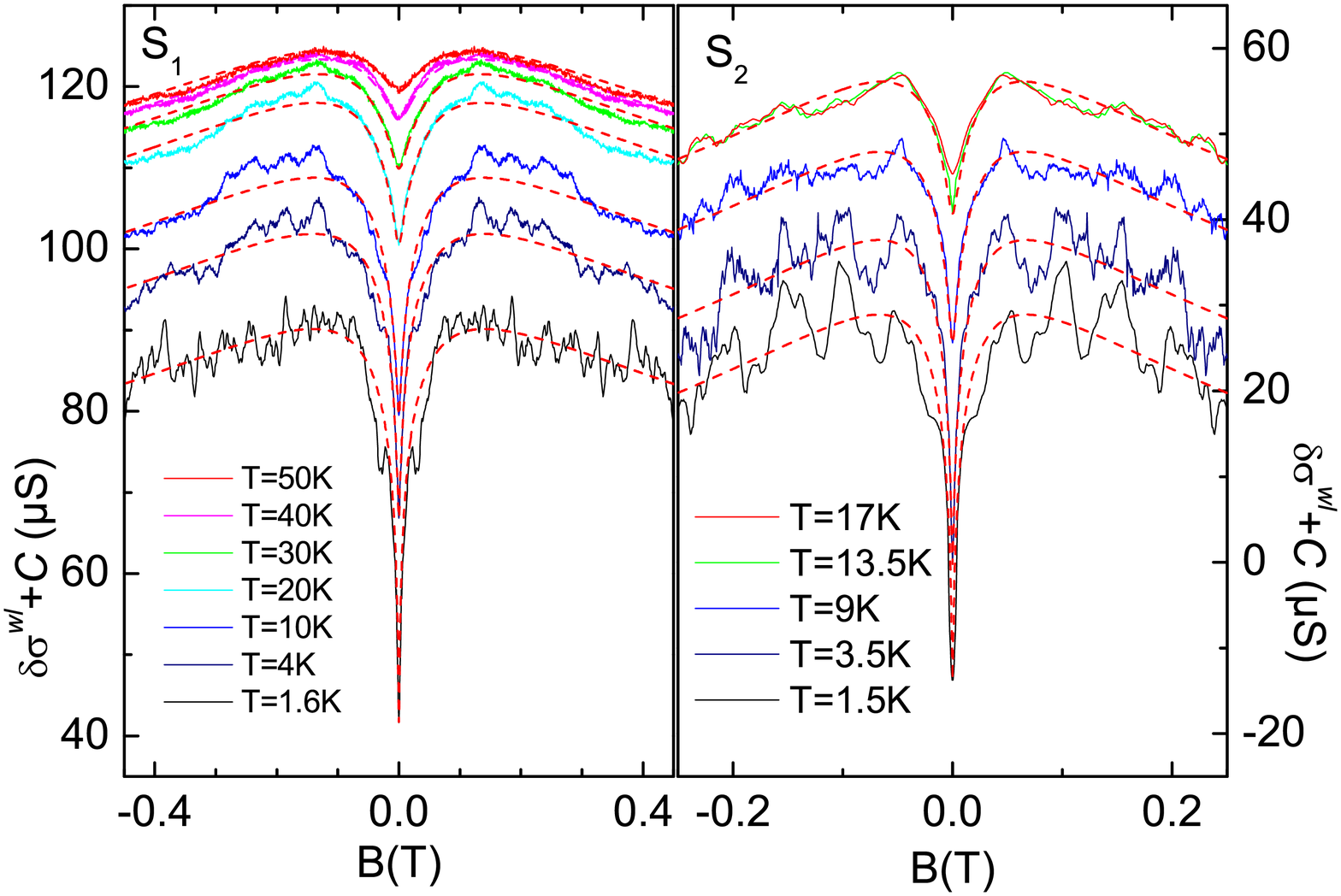}
\caption{(color online) (a) Corrections to the conductivities $\dswl+ C$ at different temperatures for sample $\Squatre$.
The $T$-dependent vertical shift $C$ has been kept for clarity.
WL fits according to Eq.~\ref{eqn:wl}  are also indicated. (b) Similar analysis for sample $\Sdeux$.}
\label{fig:wl}
\end{figure}
\begin{figure}
\includegraphics[width=0.9 \columnwidth]{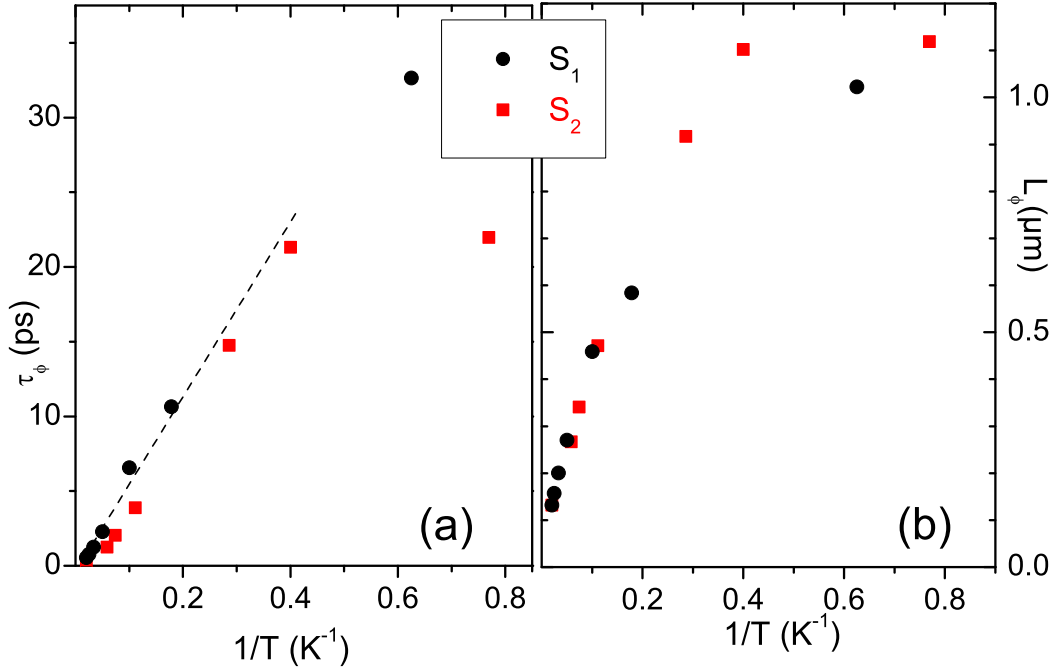}
\caption{(color online) (a) phase coherence time $\tau_\varphi$ extracted from the fit of the WL, see Fig.~\ref{fig:wl}. Black circles: sample $\Squatre$, red squares: sample $\Sdeux$. $\tau_\varphi$ is roughly proportional to $1/T$  between 60K and 2.5 K and saturates at lower temperatures. The dashed line is a guide for eyes. (b) Coherence length $L_\varphi$ {\it versus} $1/T$, for the two samples. 
 }
\label{fig:tauphi}
\end{figure}
In the diffusive regime in graphene, $\dswl$ is given by~\cite{kechedzhi2007}:
\begin{eqnarray}
\label{eqn:wl}
\dswl = 
\frac{e^2}{\pi h} \left[ 
F\left(\frac{B}{B_\varphi}\right) \right. \nonumber \\ 
\left. -F\left(\frac{B}{B_\varphi+2B_{sr}}\right)
-2F\left(\frac{B}{B_\varphi+B_{sr}+B_{lr}}\right)
 \right],
\end{eqnarray}
where $F(z)= \ln z + \Psi(\frac{1}{2}+\frac{1}{z})$, 
$\Psi$ is the digamma function,
$B_{\varphi,sr,lr}= \hbar  / 4De \tau_{\varphi,sr,lr}$ and
$\tau_{\varphi,sr,lr}$ are the coherence time, the intervalley scattering time and 
the intravalley scattering time respectively. 
For simplicity, we identify intervalley to short range (sr) scattering and intravalley  to long-range (lr) scattering.  
We take constant scattering times $\tau_{sr}$ and $\tau_{lr}$ over the whole temperature range 
and only $\tau_\varphi$ is allowed to depend on $T$.
Neglecting the warping~\cite{chen2010}, we also impose $\tau_{tr}^{-1}=  \tau_{sr}^{-1}+ \tau_{lr}^{-1}$.
We then fit the conductivities at different temperatures by Eq.~\ref{eqn:wl}.
Results are indicated by dashed lines in Fig.~\ref{fig:wl} 
(for sake of clarity, fits at high temperatures have not been reported).
Best fit is obtained with the values reported in Table I. 
We estimate the short-range scattering length $L^{sr}= \sqrt{D \tau^{sr}}$ $\approx$ 100-140 nm.  
The long-range scattering length $L^{lr}= \sqrt{D \tau^{sr}}$ $\approx$ 70-90 nm is even shorter.
These lengths are comparable to what is found in exfoliated graphene on SiO$_2$ substrate~\cite{tikho2008}.
  They are also comparable to the distance between the SiC steps below the graphene layer.
It is often stated that exfoliated graphene is much more disordered at the edge of the SiC steps~\cite{anisotropy}.
Therefore step edges could be the main source of scatterings in these samples.
 
The phase coherence time $\tau_\varphi$ obtained from the fit is plotted in Fig.~\ref{fig:tauphi}.
Apart from a saturation at low $T$, $\tau_{\varphi}$ is roughly proportional to $1/T$ between 10K-50K 
with a slope equal to $\approx$ 50 ps K  for samples $\Squatre$ and $\Sdeux$. 
We conclude that $\tau_{\varphi}$ obeys the usual
temperature dependence for electron electron scattering in the diffusive regime:
\begin{equation}
\tau_{\varphi}^{-1}= 
\beta k_B T \ln g /\hbar g,
\end{equation}
where $g$ is the reduced conductivity: $g= \sigma_0 h/e^2$ 
and the empirical coefficient $\beta$ is 1 for $\Squatre$ and 1.4 for $\Sdeux$.
Similar observations have already been done both for epitaxial~\cite{berger2007} and exfoliated graphene~\cite{tikho2008,morozov2006}.
The coherence length $L_\varphi= \sqrt{D \tau_\varphi}$ is readily calculated and plotted in fig.~4b.
At low temperature, $L_\varphi$ slightly exceeds 1$\mu$m. 
 \begin{figure}
\includegraphics[width=0.9 \columnwidth]{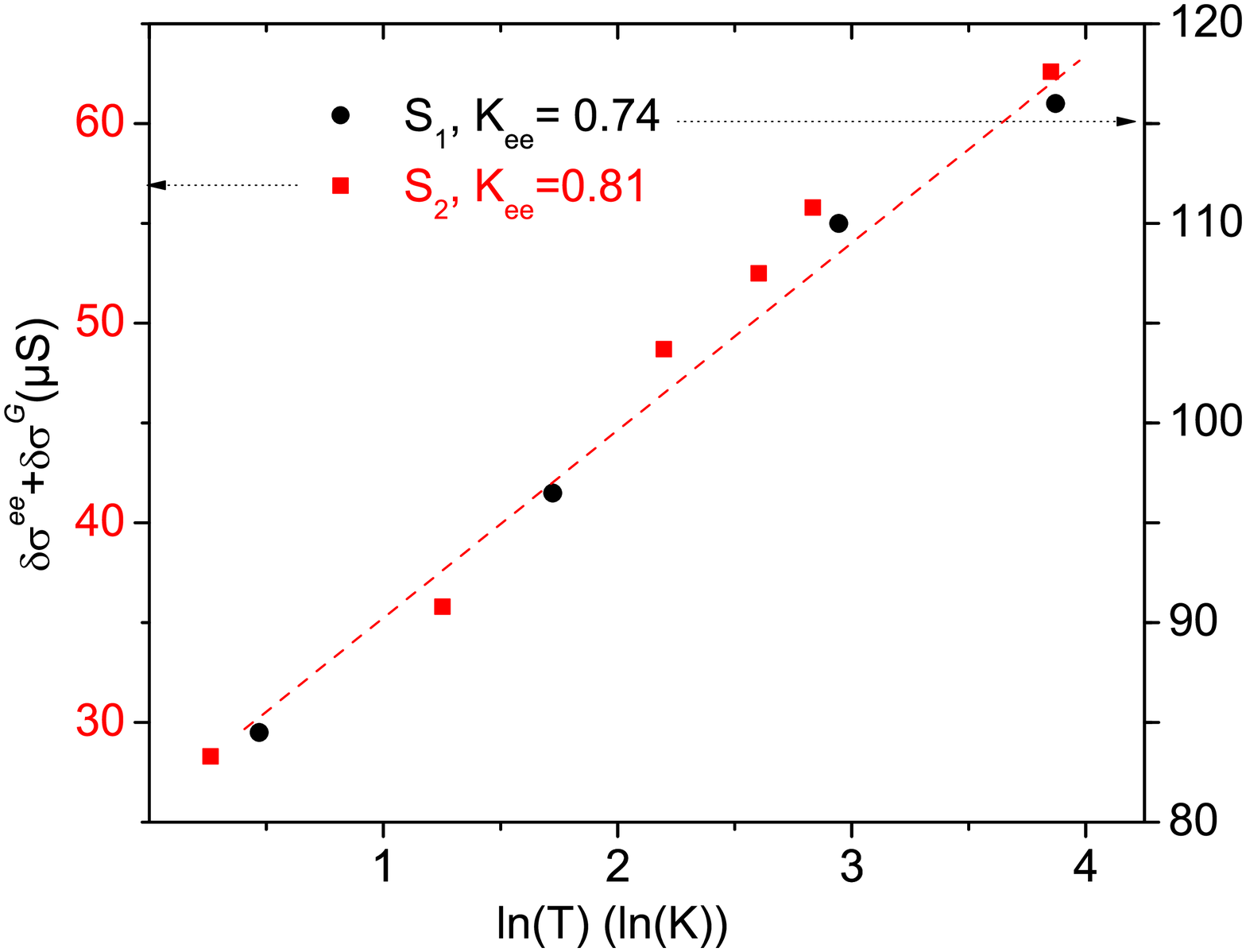}
\caption{(color online) experimental temperature dependence of 
     $C(T)= \dsgeo + \dsee(T)$ {\it vs} $\log(T)$ for samples $\Squatre$ and $\Sdeux$. The slope gives $K_{ee}=$ 0.81 and 0.74 for $\Sdeux$ and $\Squatre$ respectively.}
\label{fig:kee}
\end{figure}

\section{Correction due to Electron Electron Interaction}

We attribute the temperature dependence of the conductivity to EEI.
The Drude mobility is constant over the whole temperature range, therefore 
there is no temperature dependence of the conductivity induced 
by electron-phonon scattering. This is not surprising, as the
temperature dependence for graphene is small below the Gr\"uneisen temperature $T_{BG}$ 
given by $T_{BG} \approx$ 54$\sqrt{n_s}$ K, where the concentration $n_s$ is in unit of 10$^{12}$cm$^{-2}$ 
~\cite{guinea2007, dassarma2008}. For our samples, $T_{BG} \approx$ 54 K, and almost all our measurements
are done below the Gr\"uneisen temperature.

As  $k_BT\tau_{tr}/\hbar \ll 1$, we are by definition in the diffusive regime, for which 
EEI theory predicts a temperature dependent correction to the conductivity given by:
\begin{equation}
\label{eq:EEI}
\Delta \sigma_{ee}= K_{ee} \frac{e^2}{\pi h} \ln \frac{k_B T \tau_{tr}}{\hbar}
\end{equation}
where $K_{ee}$ is a prefactor whose value depends on the different channels contributing to the EEI~\cite{savchenko2010}. 
In fig.~\ref{fig:kee}, we show 
$\delta\sigma_{xx}(T)$
 for samples $\Sdeux$ and $\Squatre$. The expected $\log(T)$ dependence is evidenced and the slope gives 
 $K_{ee}$= 0.81$\pm 0.05$ and 0.74 $\pm 0.07$ for samples $\Sdeux$ and $\Squatre$ respectively.
These values are very similar to recent experimental findings on exfoliated graphene~\cite{savchenko2010,moser2010}.
 In a  2D system with a single valley, $K_{ee}$ takes the form~\cite{klimov2008}
\begin{equation}
\label{eqn:F0}
K_{ee}= 1+c(1- \ln(1+F_0^\sigma)/F_0^\sigma)
\end{equation}
where the unity represents the so-called 'charge' contribution (the Fock and singlet part of Hartree term), 
the $c$ prefactor is the number of 'triplet' channels (from the Hartree term)
and $F_0^\sigma$ is the liquid Fermi constant.
At low temperatures $k_B T \le \hbar/\tau_{sr}$ ($T \le$ 15 K for $\Sdeux$ and $30K$ for $\Squatre$), 
when long range and short range scattering rates are important,
the usual single-valley case is recovered with a prefactor $c=3$. 
This leads to $F_0^\sigma$ = -0.13 $\pm 0.04$.
This value is small and very close to the value of $F_0^\sigma$ recently found for exfoliated
graphene~\cite{savchenko2010}. This is somehow surprising, as the dielectric constant in SiC (10) 
is larger than in SiO$_2$ (3.9) and we would expect even smaller electron-electron interactions because of the screening of the substrate. However, numerical estimation of $F_0^\sigma$ following, for instance, Ref.~\cite{savchenko2010} gives $F_0^\sigma \approx -0.09$: a value compatible with our experiment. Also, Ref.~\cite{polini2007} predicts values for the Fermi liquid constant only slightly smaller than our findings. 
Interestingly, it was also recently quoted~\cite{savchenko2010} that 
at intermediate temperatures $\hbar/\tau_{sr} \le k_B T \le \hbar/\tau_{lr}$, 
additional triplet channels originating from pseudospin conservation (ie valley degeneracy) become relevant and 
the number of triplet channels increases to $c=7$.
However, we do not observe any change in the slope in Fig. 5 from  $T$=1.5K up to 50K. 
The situation is similar in Ref.~\cite{moser2010}, 
where EEI is observed on a temperature range on which the slope should vary, 
but where experimentally the slope remains constant.

To conclude, we show, beyond classical geometric corrections, 
weak localization, weak antilocalization and electron-electron interactions 
in epitaxial graphene. 
Weak antilocalization is observable directly in the resistances, and its analysis gives access to 
the different scattering times, which  are very close to those for exfoliated graphene on SiO$_2$ substrates.
Electron-electron interaction gives also a small correction to the conductivity, which is not significantly smaller 
than in exfoliated graphene. 

We acknowledge the EC for partial support through the RTN ManSiC Project, the French ANR for partial support through the Project Blanc GraphSiC and the Spanish Government for a grant Juan de la Cierva.  N. C. also acknowledges A. Bachtold's, A. Barreiro and J. Moser from ICN Barcelona, for technical and theoretical supports.  

begin{thebibliography}{22}
\expandafter\ifx\csname natexlab\endcsname\relax\def\natexlab#1{#1}\fi
\expandafter\ifx\csname bibnamefont\endcsname\relax
  \def\bibnamefont#1{#1}\fi
\expandafter\ifx\csname bibfnamefont\endcsname\relax
  \def\bibfnamefont#1{#1}\fi
\expandafter\ifx\csname citenamefont\endcsname\relax
  \def\citenamefont#1{#1}\fi
\expandafter\ifx\csname url\endcsname\relax
  \def\url#1{\texttt{#1}}\fi
\expandafter\ifx\csname urlprefix\endcsname\relax\def\urlprefix{URL }\fi
\providecommand{\bibinfo}[2]{#2}
\providecommand{\eprint}[2][]{\url{#2}}

\section{Appendix A}
\begin{figure}
\includegraphics[width=0.5 \columnwidth]{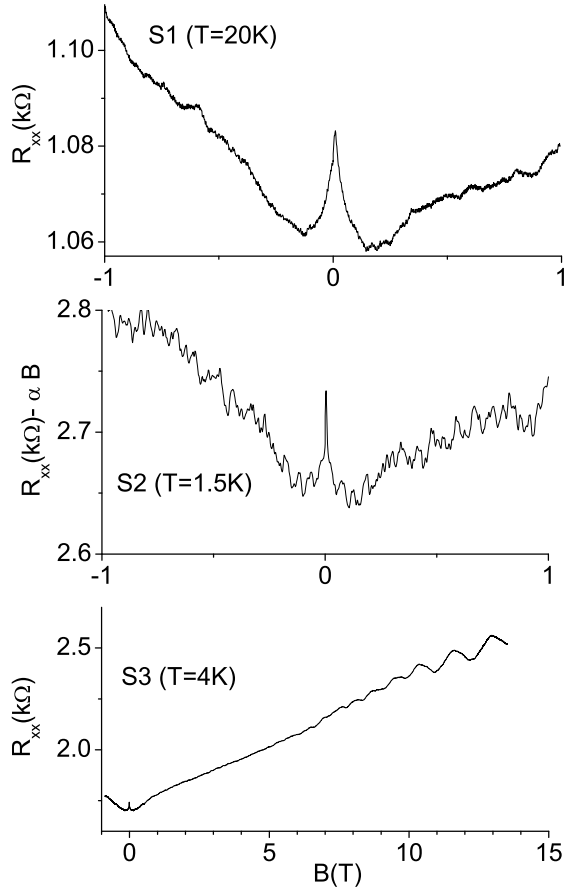}
\caption{transverse magnetoresistance for 3 different samples. A small depression attributed to WAL  is visible around the WL peak for all three samples.}
\label{fig:wal}
\end{figure}

In Fig.~\ref{fig:wal}, we present magnetoresistances of samples $S1$, $S2$ and of an additional monolayer
$S3$, more doped. A straight line $\alpha B$ has been subtracted for sample $S2$, for clarity. 
For all three samples, a dip is observed on the two sides of the central WL peak. We attribute the dip to WAL.

\section{Appendix B}
\begin{figure}
\includegraphics[width=0.6 \columnwidth]{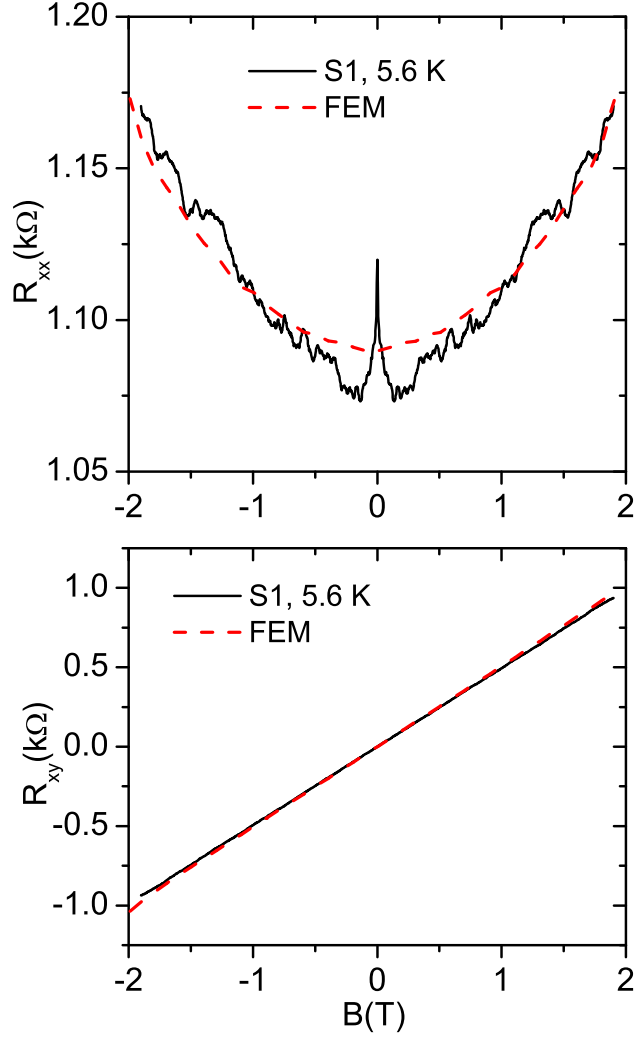}
\caption{(color online) experimental longitudinal and transverse magnetoresistances (only symmetric or antisymmetric parts have been kept for clarity) for sample $\Squatre$ are compared to the result of a numerical model based on FEM. The model imposes an increased holes concentration ($n_s= 7\cdot 10^{12}$cm$^{-2}$) and a reduced mobility ($\mu=0.06$ m$^2$V$^{-1}$s$^{-1}$) under the lateral probes. The rest of the graphene layer keeps its parameters as defined in table~\ref{table1}. }
\label{fig:simul}
\end{figure}

We suggest that most part of the parasitic magnetoresistance comes from the invasive lateral probes.
In order to sustain this assumption, classical magnetoresistances can be calculated by means of finite elements method (FEM) for the geometry of the devices. The simplest model only assumes different concentration and mobility under the lateral probes, and the results of this model are shown in Fig.~\ref{fig:simul} for sample $\Squatre$. There is a good agreement with the experiment for a reduced mobility and an increased concentration under the probes, which seems reasonable if these region have been damaged during the electron beam lithography.  
\end{document}